\documentclass[11pt]{article}
\pdfoutput=1
\usepackage{hyperref}
\usepackage{graphicx}
\usepackage{epsfig}
\usepackage{epsf}
\usepackage{epstopdf}
\usepackage{amsmath}  

\begin{document}

\def\partder#1#2{{\partial #1\over\partial #2}}

\newcommand{\be}{\begin{equation}}
\newcommand{\ee}{\end{equation}}
\newcommand{\bear}{\begin{eqnarray}}
\newcommand{\ear}{\end{eqnarray}}
\newcommand{\benn}{\begin{enumerate}}
\newcommand{\enn}{\end{enumerate}}
\newcommand{\veject}{\vfill\eject}
\newcommand{\ven}{\vfill\eject}
%
%
%

\def\ge{\hbox{$\gamma_1$}}
\def\gz{\hbox{$\gamma_2$}}
\def\gd{\hbox{$\gamma_3$}}
\def\go{\hbox{$\gamma_1$}}
\def\gt{\hbox{\$\gamma_2$}}
\def\gth{\hbox{$\gamma_3$}} 
\def\gf{\hbox{$\gamma_5\;$}}
\newcommand{\cL}{\cal L}
\newcommand{\D}{\cal D}
\newcommand{\Dhalf}{{D\over 2}}
\def\eps{\epsilon}
\def\epshalf{{\epsilon\over 2}}
\def\lag{( -\partial^2 + V)}
\def\freeexp{{\rm e}^{-\int_0^Td\tau {1\over 4}\dot x^2}}
\def\kinb{{1\over 4}\dot x^2}
\def\kinf{{1\over 2}\psi\dot\psi}
\def\expk{{\rm exp}\biggl[\,\sum_{i<j=1}^4 G_{Bij}k_i\cdot k_j\biggr]}
\def\expp{{\rm exp}\biggl[\,\sum_{i<j=1}^4 G_{Bij}p_i\cdot p_j\biggr]}
\def\expshort{{\e}^{\half G_{Bij}k_i\cdot k_j}}
\def\Dab{{(x_a-x_b)}}
\def\Dsq{{({(x_a-x_b)}^2)}}
\def\PITD{{(4\pi T)}^{-{D\over 2}}}
\def\4piTD{{(4\pi T)}^{-{D\over 2}}}
\def\4piT4{{(4\pi T)}^{-2}}
\def\TintmD{{\dps\int_{0}^{\infty}}{dT\over T}\,e^{-m^2T}
    {(4\pi T)}^{-{D\over 2}}}
\def\Tintm4{{\dps\int_{0}^{\infty}}{dT\over T}\,e^{-m^2T}
    {(4\pi T)}^{-2}}
\def\Tintm{{\dps\int_{0}^{\infty}}{dT\over T}\,e^{-m^2T}}
\def\Tint{{\dps\int_{0}^{\infty}}{dT\over T}}
\def\np{n_{+}}
\def\nm{n_{-}}
\def\Np{N_{+}}
\def\Nm{N_{-}}
\newcommand{\slG}{{{\dot G}\!\!\!\! \raise.15ex\hbox {/}}}
\newcommand{\Gd}{{\dot G}}
\newcommand{\Gund}{{\underline{\dot G}}}
\newcommand{\Gdd}{{\ddot G}}
\def\GBd12{{\dot G}_{B12}}
\def\Dx{\dps\int{\cal D}x}
\def\Dy{\dps\int{\cal D}y}
\newcommand{\1}{{\'\i}}
\def\dps{\displaystyle}
\def\sy{\scriptscriptstyle}
\def\sy{\scriptscriptstyle}

%

\newcommand{\bea}{\begin{eqnarray}}  
\newcommand{\eea}{\end{eqnarray}}  
\def\eqa{&=&}  
  
\def\a{\alpha}
\def\b{\beta}
\def\m{\mu}
\def\n{\nu}
\def\r{\rho}
\def\s{\sigma}
\def\ep{\epsilon}

\def\cosech{\rm cosech}
\def\sech{\rm sech}
\def\coth{\rm coth}
\def\tanh{\rm tanh}

  
\def\appendix{\par\clearpage
  \setcounter{section}{0}
  \setcounter{subsection}{0}
  \def\@sectname{Appendix~}
  \def\theequation{\thesection\arabic{equation}}
  \def\thesection{\Alph{section}}}
  
%
%
\def\eg{\hbox{\it e.g.}}        \def\cf{\hbox{\it cf.}}
\def\etal{\hbox{\it et al.}}
\def\dash{\hbox{---}}
%
%
%
\renewcommand{\theequation}{\arabic{section}.\arabic{equation}}
\renewcommand{\arraystretch}{2.5}
\def\R{1\!\!{\rm R}}
\def\Eins{\mathord{1\hskip -1.5pt
\vrule width .5pt height 7.75pt depth -.2pt \hskip -1.2pt
\vrule width 2.5pt height .3pt depth -.05pt \hskip 1.5pt}}
\newcommand{\symb}{\mbox{symb}}
\renewcommand{\arraystretch}{2.5}
\def\GBd12{{\dot G}_{B12}}
\def\mneg{\!\!\!\!\!\!\!\!\!\!}
\def\Mneg{\!\!\!\!\!\!\!\!\!\!\!\!\!\!\!\!\!\!\!\!}
\def\beqn*{\begin{eqnarray*}}
\def\eqn*{\end{eqnarray*}}
\def\sy{\scriptscriptstyle}
\def\footstrut{\baselineskip 12pt}
\def\square{\kern1pt\vbox{\hrule height 1.2pt\hbox{\vrule width 1.2pt
   \hskip 3pt\vbox{\vskip 6pt}\hskip 3pt\vrule width 0.6pt}
   \hrule height 0.6pt}\kern1pt}
\def\np{n_{+}}
\def\nm{n_{-}}
\def\Np{N_{+}}
\def\Nm{N_{-}}
\def\half{{1\over 2}}
\def\third{{1\over3}}
\def\fourth{{1\over4}}
\def\fifth{{1\over5}}
\def\sixth{{1\over6}}
\def\seventh{{1\over7}}
\def\eigth{{1\over8}}
\def\ninth{{1\over9}}
\def\tenth{{1\over10}}
\def\pa{\partial}
\def\ddtau{{d\over d\tau}}
\def\ge{\hbox{\textfont1=\tame $\gamma_1$}}
\def\gz{\hbox{\textfont1=\tame $\gamma_2$}}
\def\gd{\hbox{\textfont1=\tame $\gamma_3$}}
\def\go{\hbox{\textfont1=\tamt $\gamma_1$}}
\def\gt{\hbox{\textfont1=\tamt $\gamma_2$}}
\def\gth{\hbox{\textfont1=\tamt $\gamma_3$}} 
\def\gf{\hbox{$\gamma_5\;$}}
\def\ie{\hbox{$\textstyle{\int_1}$}}
\def\iz{\hbox{$\textstyle{\int_2}$}}
\def\id{\hbox{$\textstyle{\int_3}$}}
\def\ldop{\hbox{$\lbrace\mskip -4.5mu\mid$}}
\def\rdop{\hbox{$\mid\mskip -4.3mu\rbrace$}}
\def\eps{\epsilon}
\def\epshalf{{\epsilon\over 2}}
\def\e{\mbox{e}}
\def\mn{{\mu\nu}}
\def\exmn{{(\mu\leftrightarrow\nu )}}
\def\ab{{\alpha\beta}}
\def\exab{{(\alpha\leftrightarrow\beta )}}
\def\g{\mbox{g}}
\def\kinb{{1\over 4}\dot x^2}
\def\kinf{{1\over 2}\psi\dot\psi}
\def\expk{{\rm exp}\biggl[\,\sum_{i<j=1}^4 G_{Bij}k_i\cdot k_j\biggr]}
\def\expp{{\rm exp}\biggl[\,\sum_{i<j=1}^4 G_{Bij}p_i\cdot p_j\biggr]}
\def\expshort{{\e}^{\half G_{Bij}k_i\cdot k_j}}
\def\expabb{{\e}^{(\cdot )}}
\def\PITD{{(4\pi T)}^{-{D\over 2}}}
\def\4piTD{{(4\pi T)}^{-{D\over 2}}}
\def\4piT4{{(4\pi T)}^{-2}}
\def\TintmD{{\dps\int_{0}^{\infty}}{dT\over T}\,e^{-m^2T}
    {(4\pi T)}^{-{D\over 2}}}
\def\Tintm4{{\dps\int_{0}^{\infty}}{dT\over T}\,e^{-m^2T}
    {(4\pi T)}^{-2}}
\def\Tintm{{\dps\int_{0}^{\infty}}{dT\over T}\,e^{-m^2T}}
\def\Tint{{\dps\int_{0}^{\infty}}{dT\over T}}
\def\pint{{\dps\int}{dp_i\over {(2\pi)}^d}}
\def\Dx{\dps\int{\cal D}x}
\def\Dy{\dps\int{\cal D}y}
\def\Dpsi{\dps\int{\cal D}\psi}
\def\Tr{{\rm Tr}\,}
\def\tr{{\rm tr}\,}
\def\sumij{\sum_{i<j}}
\def\freeexp{{\rm e}^{-\int_0^Td\tau {1\over 4}\dot x^2}}
\def\arraystretch{2.5}
\def\Ge{\mbox{GeV}}
\def\dA{\partial^2}
\def\DA{\sqsubset\!\!\!\!\sqsupset}
\def\FFdual{F\cdot\tilde F}
\def\mn{{\mu\nu}}
\def\rs{{\rho\sigma}}
\def\oplusotimes{{{\lower 15pt\hbox{$\scriptscriptstyle \oplus$}}\atop{\otimes}}}
\def\perppar{{{\lower 15pt\hbox{$\scriptscriptstyle \perp$}}\atop{\parallel}}}
\def\oopp{{{\lower 15pt\hbox{$\scriptscriptstyle \oplus$}}\atop{\otimes}}\!{{\lower 15pt\hbox{$\scriptscriptstyle \perp$}}\atop{\parallel}}}
%
%
\def\bbbr{{\rm I\!R}}
\def\bbbone{{\mathchoice {\rm 1\mskip-4mu l} {\rm 1\mskip-4mu l}
{\rm 1\mskip-4.5mu l} {\rm 1\mskip-5mu l}}}
\def\bbbz{{\mathchoice {\hbox{$\sf\textstyle Z\kern-0.4em Z$}}
{\hbox{$\sf\textstyle Z\kern-0.4em Z$}}
{\hbox{$\sf\scriptstyle Z\kern-0.3em Z$}}
{\hbox{$\sf\scriptscriptstyle Z\kern-0.2em Z$}}}}

\renewcommand{\thefootnote}{\protect\arabic{footnote}}
%

\begin{center}
{\huge\bf Asymptotic behaviour of the QED perturbation series
\footnote{To appear in the proceedings of ``5th Winter Workshop on Non-Perturbative Quantum Field Theory'', 22-24 March 2017, Sophia-Antipolis, France.}}
\vspace{5pt}

\vskip.7cm

{\large Idrish Huet$^{a}$, Michel Rausch de Traubenberg$^{b}$, 
\underline{Christian Schubert}$^{c}$}
\\[1.5ex]

\begin{itemize}
\item [$^a$]
{\it 
Facultad de Ciencias en  F\'{\i}sica y Matem\'aticas, Universidad Aut\'onoma de
Chiapas, Ciudad Universitaria, Tuxtla Guti\'errez 29050, M\'exico.\\
idrish@ifm.umich.mx
}
\item [$^b$]
{\it
IPHC-DRS, UdS, IN2P3,\\
23 rue du Loess, F-67037 Strasbourg Cedex, France.\\
Michel.Rausch@IReS.in2p3.fr
}

\item [$^c$]
{\it 
Instituto de F\'{\i}sica y Matem\'aticas,
\\
Universidad Michoacana de San Nicol\'as de Hidalgo,\\
Edificio C-3, Apdo. Postal 2-82,\\
C.P. 58040, Morelia, Michoac\'an, M\'exico.\\
schubert@ifm.umich.mx
}
\end{itemize}
\end{center}
\vspace{1cm}
 {\large \bf Abstract:}
\begin{quotation}
In this talk, I will summarize the present state of a long-term effort to
obtain information on the high-order asymptotic behaviour of the QED
perturbation series through the effective action. Starting with the
constant-field case,  I will discuss the Euler-Heisenberg Lagrangian in
various dimensions, and  up to the three-loop level. This Lagrangian holds
the information on the  N-photon amplitudes in the low-energy limit, and
combining it with spinor helicity methods explicit all-N results can be
obtained at the one-loop and, for the ``all + '' amplitudes, also at the two-loop
level. For the imaginary part of the Euler-Heisenberg  Lagrangian, an
all-loop formula has been conjectured independently by Affleck, Alvarez and Manton for Scalar
QED, and by Lebedev and Ritus for Spinor QED. 
This formula can be related  through a Borel
dispersion relation to the leading large-N behaviour of  the N-photon
amplitudes. It is analytic in the fine structure constant,  which is
puzzling and suggests a diagrammatic investigation of the large-N limit
in perturbation theory. Preliminary results of such a study for
the 1+1 dimensional case throw doubt on the validity
of the conjecture.

\end{quotation}
\vfill\eject
\pagestyle{plain}
\setcounter{page}{1}
\setcounter{footnote}{0}

\vspace{10pt}
\section{Motivation}
\label{motivation}
\renewcommand{\theequation}{1.\arabic{equation}}
\setcounter{equation}{0}

In 1952 Dyson \cite{dyson52} shocked the high energy physics community by declaring that, quite generally,
the QED perturbation series cannot converge. Writing the series as

\bear
F(e^2) = c_0 + c_2 e^2 + c_4 e^4 + \ldots  \, ,
\ear
Dyson argues: ``Suppose, if possible, that the series converges for some positive value of $e^2$; 
this implies that $F(e^2)$ is an analytic function of $e$ at $e=0$. Then for sufficiently small values of $e$, $F(-e^2)$ will also be a well-behaved analytic function with a convergent power-series expansion''. 

He then argues that, on physical grounds, this cannot be the case,
since for $e^2<0$ the QED vacuum
will be unstable due to a runaway production of $e^+e^-$ pairs which coalesce into
like-charge groups. 

Only shortly later  C.A. Hurst \cite{hurst} already provided a mathematical proof
of this fact for scalar $\lambda \phi^3$ theory. 
The proof is essentially based on the following three elements:

\begin{enumerate}

\item
The use of the inequality

\bear
\prod_{i=1}^F \biggl({1\over p_i^2 + \kappa^2}\biggr)
\geq
{F^F\over \Bigl(\sum_{i=1}^Fp_i^2 + F\kappa^2\Bigr)^F}
\ear
to establish lower bounds for arbitrary Feynman diagrams (in the Euclidean). 

\item
Proof that the number of distinct Feynman
diagrams at $n^{\rm th}$ loop order grows like $({n\over 2})!n!$ 

\item
Absence of sign cancellations between graphs. 

\end{enumerate}


In 1979, `t Hooft \cite{thooft-renormalons} found another very general, but very different, argument against 
convergence of the perturbation series based on renormalon chains. 
Thus today it is believed that  the perturbation series in nontrivial quantum field theories generically 
is asymptotical rather than convergent, so that summation methods must be used.
Of those by far the most important one is Borel summation, since it is ideally suited to the typical
factorial growth of perturbation theory coefficients. Let me remind you that, for a factorially divergent series,  

\bear
F(g)  \sim \sum_{n=0}^{\infty} c_n g^{n+1}
\ear
one defines the {Borel transform} as

\bear
B(t) \equiv \sum_{n=0}^{\infty} c_n {t^n\over n!} \, .
\ear
If $B(t)$ has no singularities on the positive real axis 
and does not increase too rapidly at infinity, 
one can also define the Borel integral

\bear
\tilde F (g) \equiv \int_0^{\infty}dt\,  e^{-{t\over g}}B(t)
\label{borelint}
\ear
$\tilde F$ is the {Borel sum} of the original series $F$. $F$ is asymptotic to $\tilde F$ by construction,
although the true physical quantity represented by the series $F$ might still differ from $\tilde F$ by
nonperturbative terms. 
The Borel transform remains a useful concept even when it leads to singularities, 
since those usually contain information on the large-order  structure of the theory.
In many cases they can be traced either to {instantons}, {renormalons}  or {Euclidean bounces}.

Until recently, there was a dearth of nontrivial examples for field theory models where sufficient information would
be available to decide the question of Borel summability in a definite manner. Fortunately, this has changed through the
advent of supersymmetry; in recent years 
Borel summability (or Borel non-summability)
has been rigorously demonstrated in a number of supersymmetric models \cite{russo-borel}.

Even when Borel summability does not apply, Borel analysis can still be very useful through the use of
Borel dispersion relations. This goes as follows. 
Assume that a function $F(g)$ has an asymptotic series expansion 

\bear
F(g) \sim \sum_{n=0}^{\infty} c_n g^n
\ear
where the expansion coefficients $c_n$ have the leading-order large $n$ behaviour

\bear
c_n \sim \rho^n \Gamma(\mu n + \nu)
\label{fit}
\ear
with some real constants $\rho >0$,  $\mu > 0 $ and $\nu$. It is easy to see that such a series
is not Borel-summable, since the Borel integral (\ref{borelint}) can never converge (for example, in the 
textbook case $\mu = \nu =1$ it has a pole at $t=1/\rho$)  . Nevertheless, applying 
a dispersion relation to this integral one can show
that the leading contribution to its imaginary part for small $g$ is given by 

\begin{eqnarray}
{\rm Im} F(g)\sim\frac{\pi}{\mu}\left(\frac{1}{\rho g} \right)^{\nu/\mu}
\exp\left[-\left(\frac{1}{\rho g}\right)^{1/\mu}\right]
\, .
\label{boreldisp}
\end{eqnarray}

Coming back to the case of QED, given the arguments by Dyson and `t Hooft it is certainly save to exclude 
a nonzero convergence radius of the full QED perturbation series. However, despite of the immense 
work that has gone into low-order perturbative QED computations, presently still little is known about the
precise large-order behavior of the coefficients. Contrary to the case of scalar field theories mentioned above, 
straightforward estimates based on lower bounds for individual diagrams cannot be used in gauge theory,
since here Feynman diagrams come with different signs, and gauge invariance is known to lead to 
cancellations between them. And these cancellations are particularly extensive in the abelian case, where
there are no obstructing color factors. Thus QED in this respect is more difficult than QCD, which is made worse by the
absence of  (spacetime) instantons in QED, which in the nonabelian case can provide some large-order information. 
In 1977 Cvitanovic \cite{cvitanovic77} suggested, based on an analysis of the calculation of the three-loop anomalous
magnetic momentum $g-2$ which he had  done with Kinoshita \cite{cvikin}, that these cancellations should be taken into account
by counting the number of classes of gauge-invariant diagrams, rather than the number of individual diagrams. 
He also conjectured that, for the case of $g-2$, they reduce the coefficients of the perturbation series sufficiently
to make it convergent in the quenched approximation. This conjecture, although nowadays forgotten,
is actually still standing, since neither Dyson's nor  `t Hooft's arguments work in the absence of fermionic bubbles.

Here I will summarize the state of a long-term effort \cite{37,51,52,56,60,66,81,85,inprep} 
to get information on the high-order behaviour of the QED perturbation series using the 
Euler-Heisenberg Lagrangian and its higher-loop radiative corrections. 

\section{The 1-loop Euler-Heisenberg Lagrangian.}
\label{1loopEHL}
\renewcommand{\theequation}{2.\arabic{equation}}
\setcounter{equation}{0}

The Euler-Heisenberg Lagrangian (``EHL'') is the one-loop QED effective Lagrangian for a constant external field.
Euler and Heisenberg \cite{eulhei} obtained for it in 1936 the following well-known proper-time representation:

\bear
{\cal L}^{(1)}_{\rm spin}(F)&=& - {1\over 8\pi^2}
\int_0^{\infty}{dT\over T^3}
\,\e^{-m^2T}
\biggl\lbrack
{(eaT)(ebT)\over {\rm tanh}(eaT)\tan(ebT)} 
- {e^2\over 3}(a^2-b^2)T^2 -1
\biggr]
\, .
\nonumber\\
\label{ehspin}
\ear
Here $a,b$ are the two
invariants of the Maxwell field, 
related to $\bf E$, $\bf B$ by 

\bear
a^2-b^2 = B^2-E^2,\quad ab = {\bf E}\cdot {\bf B} \, .
\label{ab}
\ear
The analogous result for Scalar QED was obtained by Weisskopf \cite{weisskopf},
but will be called ``Scalar Euler-Heisenberg Lagrangian'' in the following.

\bear
{\cal L}_{\rm scal}^{(1)}(F)&=&  {1\over 16\pi^2}
\int_0^{\infty}{dT\over T^3}
\,\e^{-m^2T}
\biggl[
{(eaT)(ebT)\over \sinh(eaT)\sin(ebT)} 
+{e^2\over 6}(a^2-b^2)T^2 -1
\biggr]
\, .
\nonumber\\
\label{ehscal}
\ear
The Euler-Heisenberg Lagrangian (``EHL'')  holds information on
the one-loop $N$ photon amplitudes, 
but only in the low energy limit (since a constant field can emit only 
zero energy photons). 

\begin{figure}[htbp]
\begin{center}
\includegraphics[scale=.6]{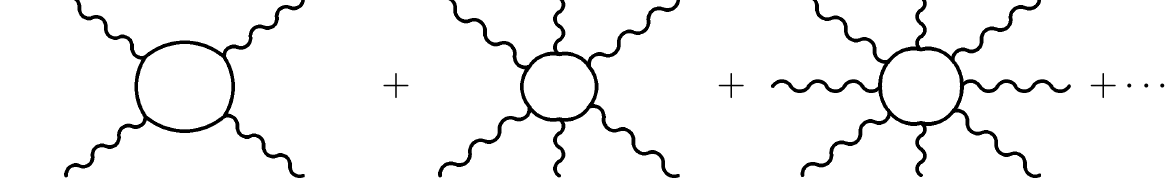}
\end{center}
\caption{Sum of diagrams equivalent to the one-loop EHL.}
\label{fig-EHL1loop}
\end{figure}

Thus diagrammatically 
${\cal L}^{(1)}(F)$ is equivalent to the sum of the Feynman graphs shown in Fig. \ref{fig-EHL1loop},
where all photon energies are small
compared to the electron mass, $\omega_i\ll m$.

\bigskip

In \cite{56} it was shown how to carry out explicitly the construction of  these limiting low-energy amplitudes 
starting from the weak field expansion of the EHL,

\bear
{\cal L} (F) = \sum_{k,l} c_{kl}\, a^{2k}b^{2l} \, .
\label{wfe}
\ear
It turned out that, if one fixes the number of photons, their momenta $k_1,\ldots,k_N$ and
a helicity assignment for each photon, then in this limit all the dependence on momenta 
and polarizations is carried by a unique invariant. Thus the magnitude of the amplitude can be specified by a
single number, which will be essential for our whole approach.

Except for the purely magnetic case, the EHL has also an imaginary part related
to vacuum pair creation by the electric field component (to be called ``Sauter-Schwinger pair
creation'' in the following) \cite{sauter,schwinger51}. 
In the purely electric case one finds, from the poles in the $T$ - integration, the
following decomposition due to Schwinger \cite{schwinger51},

\begin{eqnarray}
{\rm Im} {\cal L}^{(1)}(E) &=&  \frac{m^4}{8\pi^3}
\beta^2\, \sum_{k=1}^\infty \frac{1}{k^2}
\,\exp\left[-\frac{\pi k}{\beta}\right] \, ,
\nonumber\\
{\rm Im}{\cal L}_{\rm scal}^{(1)}(E) 
&=&
-\frac{m^4}{16\pi^3}
\beta^2\, \sum_{k=1}^\infty \frac{(-1)^{k}}{k^2}
\,\exp\left[-\frac{\pi k}{\beta}\right] 
\nonumber\\
\label{schwinger}
\end{eqnarray}
($\beta = eE/m^2$). 
In the following we will focus on the weak field limit  $\beta \ll 1$ where only the first 
of these ``Schwinger exponentials'' is relevant.

The nonperturbative dependence of the Schwinger exponentials on the field supports the interpretation
of field-induced pair creation as a vacuum tunneling effect, as proposed by Sauter
as early as 1931 \cite{sauter}.

\begin{figure}[ht]
\centerline{\hspace{-30pt}\includegraphics[scale=.6]{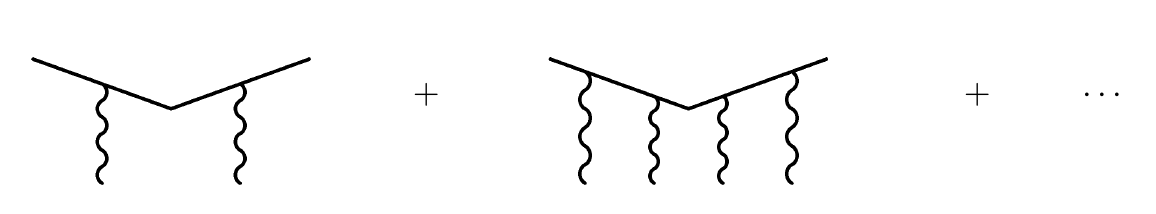}}
\caption{Cut diagrams giving the imaginary part of the $N$ - photon amplitudes.}
\label{fig-schwinger_tree_diags}
\end{figure}

%
As usual in quantum field theory, the real and imaginary parts of the EHL are related by a
dispersion relation. For the $N$ - photon amplitudes at full momentum, this would
be a standard dispersion relation performed diagram-by-diagram, relating the diagrams
of Fig. \ref{fig-EHL1loop} to the ``cut diagrams'' shown in Fig. \ref{fig-schwinger_tree_diags},
involving on-shell electrons. 

However, in the zero-energy limit the cut diagrams all vanish, since 
a finite number of zero-energy photons cannot create a pair on-shell.
Thus what counts here is only the asymptotic behavior for a large number of photons, and 
instead of an ordinary dispersion relation we have to use a {\it Borel dispersion relation}.
This works in the following way \cite{37}. 

Consider the purely magnetic EHL. Expanding it out in powers of the field yields

\bear
{\cal L}^{(1)}(B)&=& - {1\over 8\pi^2}
\int_0^{\infty}{dT\over T^3}
\,\e^{-m^2T} 
\Biggl[
{eBT\over \tanh(eBT)} - {1\over 3}(eBT)^2 -1
\Biggr]
\nonumber\\
&=&
\frac{2m^4}{\pi^2}
\sum_{n=2}^{\infty}
c_n^{(1)}g^n
\label{EHLmag}
\ear
with an effective expansion parameter $g=\Bigl({eB\over m^2}\Bigr)^2$,
and coefficients $c_n^{(1)}$ that can be written in terms of the Bernoulli
numbers $B_n$: 

\bear
c_n^{(1)} = -{2^{2n-4}B_{2n}\over (2n)(2n-1)(2n-2)} \, .
\label{cnBernoulli}
\ear
Here $c_n^{(1)}$ holds information on the $N=2n$ photon amplitudes.  
The asymptotic behavior of the coefficients can be easily studied
using well-known properties of the Bernoulli numbers. One finds 

\bear
c_n^{(1)} \quad
{\stackrel {n\to\infty} \sim} \quad
(-1)^n{1\over 8}{\Gamma(2n-2)\over\pi^{2n}}
\biggl(1+{1\over 2^{2n}}+{1\over 3^{2n}}+ \ldots \biggr)
\label{cnexpand}
\ear
Thanks to the factor $(-1)^n$, the individual terms on the right hand side of (\ref{cnexpand})
all give convergent Borel integrals. This is one (rather roundabout) way of
seeing that the purely magnetic EHL has no imaginary part, and does not pair create. 

The analogous expansion for the purely electric field case is almost the same:

\bear
{\cal L}^{(1)}(E)&=& - {1\over 8\pi^2}
\int_0^{\infty}{dT\over T^3}
\,\e^{-m^2T} 
\Biggl[
{eET\over \tan(eET)} + {1\over 3}(eET)^2 -1
\Biggr]
\nonumber\\
&=&
\frac{2m^4}{\pi^2}
\sum_{n=2}^{\infty}
(-1)^n c_n^{(1)}g^n
\label{EHLel}
\ear
where now  $g=\Bigl({eE\over m^2}\Bigr)^2$, but with the same $c_n^{(1)}$. 
However, the additional factor $(-1)^n$ makes the series non-alternating,
which is crucial, because now the termwise use of the expansion (\ref{cnexpand})
leads to divergent Borel integrals. These divergent integrals do, however,
all possess well-defined imaginary parts, by a (now ordinary) dispersion relation.
One finds a perfect match between the expansion (\ref{cnexpand}) and 
Schwinger's expansion (\ref{schwinger}):

\bear
c_n^{(1)} \sim (-1)^n{1\over 8}{\Gamma(2n-2)\over\pi^{2n}}
&\rightarrow&
{\rm Im}\,{\cal L}^{(1)}(E) \sim {m^4\over 8\pi^3}
\biggl({eE\over m^2}\biggr)^2{\rm exp}\biggl(-{\pi m^2\over eE}\biggr) \, ,
\nonumber\\
c_n^{(1)} \sim (-1)^n{1\over 8}{\Gamma(2n-2)\over\pi^{2n}}
{1\over 2^{2n}}
&\rightarrow&
{\rm Im}\,{\cal L}^{(1)}(E) \sim {m^4\over 8\pi^3}
\biggl({eE\over m^2}\biggr)^2{1\over 2^2}{\rm exp}\biggl(-2{\pi m^2\over eE}\biggr) \, ,
\nonumber\\
\vdots && \vdots
\nonumber\\
\label{match}
\ear

\section{The Euler-Heisenberg Lagrangian at higher loops.}
\renewcommand{\theequation}{3.\arabic{equation}}
\setcounter{equation}{0}

\begin{figure}[h]
{\centering
\hspace{70pt}\includegraphics[scale=.7]{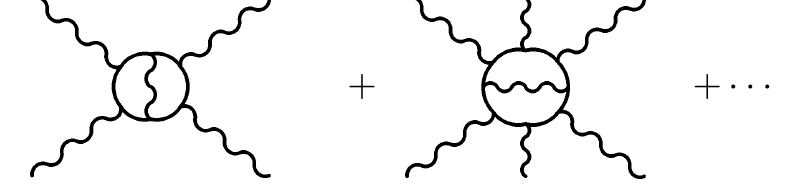}
}
\caption{Feynman diagrams contributing to the 2-loop EHL.}
\label{fig-2loopEHL}
\end{figure}

Proceeding to higher loop orders, the two-loop EHL ${\cal L}^{(2)}$  is generated by 
the diagrams shown in Fig. \ref{fig-2loopEHL} (here and in the following it is understood 
that internal photon corrections are  put in all possible ways). 

\bigskip

\begin{figure}[h]
\hspace{50pt}
{\centering
\includegraphics[scale=.7]{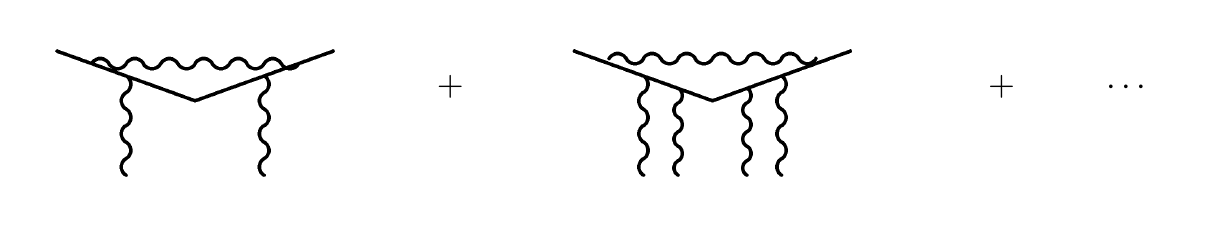}
}
\caption{Feynman diagrams contributing to 2-loop Schwinger pair creation.}
\label{fig-2loopSchwinger}
\end{figure}

As in the one-loop case, ${\cal L}^{(2)}$ will have an imaginary part iff the field is not
purely magnetic. This imaginary part corresponds to the one-loop ``cut diagrams''  
depicted in Fig. \ref{fig-2loopSchwinger}.

The two-loop EHL was first studied by V.I. Ritus, both for Spinor \cite{ritusspin} and Scalar QED  \cite{ritusscal}.
These calculations, as well as later recalculations \cite{ditreubook,18,66},
resulted in a type of rather intractable two-parameter integrals.
However, the first few coefficients of the weak-field expansions of the two-loop
EHLs have been computed \cite{18,37,66}. 
As to the imaginary parts, the Schwinger formulas (\ref{schwinger}) generalize
to the two-loop level as follows \cite{lebrit}:

\begin{eqnarray}
{\rm Im} {\cal L}^{(2)} (E) &=&  \frac{m^4}{8\pi^3}
\beta^2\,
\sum_{k=1}^\infty
\alpha\pi K_k^{\rm spin}(\beta)
\,\exp\left[-\frac{\pi k}{\beta}\right] \, ,
\nonumber\\
{\rm Im} {\cal L}_{\rm scal}^{(2)} (E) &=&  \frac{m^4}{16\pi^3}
\beta^2\,
\sum_{k=1}^\infty
(-1)^{k+1}
\alpha\pi K_k^{\rm scal}(\beta)
\,\exp\left[-\frac{\pi k}{\beta}\right] \, .
\nonumber\\
\label{ImL2}
\end{eqnarray}
($\alpha=\frac{e^2}{4\pi}$), where

\begin{eqnarray}
K_k^{\rm scal,spin}(\beta) &=& -{c_k\over \sqrt{\beta}} + 1 + {\rm O}(\sqrt{\beta}) \, ,
\nonumber\\
c_1 = 0,\quad && \quad
c_k = {1\over 2\sqrt{k}}
\sum_{l=1}^{k-1} {1\over \sqrt{l(k-l)}},
\quad k \geq 2 \, ,
\label{expK}
\end{eqnarray}
(these coefficients, also called $c_n$, should not be confused with the weak-field expansion coefficients
introduced above).
Thus at two-loop the $k$th Schwinger-exponential appears with a prefactor which
is still a function of the field strength, and of which presently only the lowest order terms in 
the weak-field expansion are known. Still, things become very simple at leading order
in this expansion: Adding the one-loop and two-loop EHL's, one finds, 
e.g. for the spinor QED case \cite{lebrit},
  
\begin{eqnarray}
{\rm Im} {\cal L}^{(1)} (E) +
{\rm Im}{\cal L}^{(2)} (E) 
\,\,\,\, {\stackrel{\beta\to 0}{\sim}} \,\,\,\,
 \frac{m^4\beta^2}{8\pi^3}
\bigl(1+\alpha\pi\bigr)
\,{\rm e}^{-{\pi\over\beta}}
\label{Im1plus2}
\end{eqnarray}
and this result is spin-independent (but for the normalization). 
In (\cite{lebrit} )it was further noted that, if one assumes that in this 
weak-field approximation higher order corrections just lead to an exponentiation,

\begin{eqnarray}
\sum_{l=1}^{\infty}
{\rm Im} {\cal L}^{(l)} (E) \,\,\,\, {\stackrel{\beta\to 0}{\sim}} \,\,\,\,
{\rm Im} {\cal L}^{(1)} (E) 
\,{\rm e}^{\alpha\pi}
\label{Imall}
\end{eqnarray} 
then the ${\rm e}^{\alpha\pi}$ factor can be absorbed into the Schwinger factor ${\rm e}^{-{\pi\over\beta}}$
by the following mass-shift,

\bear
m(E) \approx m - \frac{\alpha}{2}\frac{eE}{m} \, .
\ear
Moreover, the existence of this mass-shift can be independently confirmed in two different ways.
First, the same mass shift had been found by Ritus already before 
in the crossed process of electron propagation in the electric field \cite{ritusmass}.
Second, in the tunneling picture it can be interpreted as the correction to the
Schwinger pair creation rate due to the pair being created with a negative Coulomb
interaction energy  at a definite distance, taking the Coulomb interaction into account at the
one-photon exchange level \cite{lebrit}.

And, although this was not known to the authors of \cite{lebrit}, an analogous exponentiation had already been
conjectured two years before for the Scalar QED case by Affleck et al. \cite{afalma}. However,
those authors used a totally different approach based on {\it worldline instantons}. To explain this concept,
we first have to discuss Feynman's {\it worldline path integral representation} of the QED effective
action.

\section{Worldline representation of the QED effective action}
\renewcommand{\theequation}{4.\arabic{equation}}
\setcounter{equation}{0}

In 1950 Feynman presented, in an appendix to one of his groundbreaking papers on the 
modern, manifestly relativistic formalism of QED \cite{feynman1950}, also
an alternative first-quantized formulation of scalar QED, ``for its own interest as an alternative
to the formulation of second quantization''. There he provides a simple rule for constructing
the complete scalar QED S-matrix by representing the scalar particles in terms
of relativistic particle path integrals, and coupling them through photons in all possible ways. 
Upon restriction to the purely photonic part of the S-matrix (no
external scalars), and to the ``quenched'' contribution (only one virtual scalar), 
this ``worldline representation'' can be stated very compactly in terms of
the (quenched) effective action $\Gamma [A]$:

\bear
\Gamma_{\rm scalar}[A] &=&
\int d^4x\, {\cal L}_{\rm scalar}[A] =
\int_0^{\infty}{dT\over T}\,{\rm e}^{-m^2T}
{\displaystyle \int_{x(T)=x(0)}}{\cal D}x(\tau)
\, e^{-S[x(\tau)]}  \, .\nonumber\\
\label{Gammascal}
\ear
Here $T$ denotes the proper-time of the scalar particle in the loop,
$m$ its mass, and $ \int_{x(T)=x(0)}{\cal D}x(\tau)$ a path integral over
all closed loops in spacetime with fixed periodicity in the proper-time.
The worldline action $S[x(\tau)]$ has three parts,  
 
\bear
S&=& S_0 + S_{\rm ext} + S_{\rm int} \, .
\label{Ssplit}
\ear
They are given by

\bear
S_0 &=& \int_0^T d\tau {\dot x^2\over 4} \hspace{158pt}  {\rm (free\,\, propagation)}\, , \nonumber\\
S_{\rm ext} &=& ie\int_0^T \dot x^{\mu}A_{\mu}(x(\tau)) 
\hspace{120pt} {\rm (external\,\, photons)}\, , \nonumber\\
S_{\rm int} &=&
-{e^2\over 8\pi^2}\int_0^Td\tau_1\int_0^Td\tau_2 {\dot x(\tau_1)\cdot\dot x(\tau_2)\over
(x(\tau_1)-x(\tau_2))^2}
\qquad{\rm (internal\,\, photons)} \, .
\nonumber\\
\label{Sparts}
\ear
The kinetic term $S_0$ describes the free propagation of the scalar, $S_{\rm ext}$
its interaction with the external field, and $S_{\rm int}$ generates the corrections due to
internal photon exchanges in the loop. Expanding out the two interaction
exponentials leads back to Feynman diagrams, 
however with the important
difference that no particular ordering of the photon legs along the loop needs
to be fixed. Thus the term $S_{\rm ext}$ alone upon expansion yields the diagrams
of Fig. \ref{fig-EHL1loop} (where each leg now stands for an interaction with the
arbitrary field $A(x)$). 

The ``worldline instanton'' of Affleck et al. \cite{afalma} is an extremal trajectory of the
worldline path integral for a stationary phase approximation.
For the case of a constant electric field in the $z$ direction this 
extremal action trajectory is given by a circle in the (euclidean) $t-z$ plane:

\bear
x_{\rm instanton}(\tau) = {m\over eE}\bigr(0,0,{\rm cos}(2\pi \tau /T),{\rm sin}(2\pi \tau /T)\bigl) \, .
\label{instanton}
\ear
It can be shown that in the weak field (= large mass) limit the imaginary (although not the
real) part of the effective Lagrangian can be well-approximated by replacing the
path integral with this single trajectory:

\bear
{\rm Im}{\cal L}_{\rm scalar}^{({\rm quenched})}(E)  \sim {\rm e}^{-S[x_{\rm instanton}]} \, .
\label{singletraj}
\ear
This is easily evaluated to be

\bear
(S_0+S_{\rm ext})[x_{\rm instanton}] = \pi {m^2\over eE}, \qquad
S_{\rm int}[x_{\rm instanton}] = - \alpha\pi \,. \nonumber\\
\ear
Thus the contribution of $S_0+S_{\rm ext}$ just reproduces the
leading (one loop) Schwinger exponential of (\ref{schwinger}), and the 
one of $S_{\rm int}$ the ${\rm e}^{\alpha\pi}$ factor. 

\begin{figure}[h]
{\centering
\includegraphics{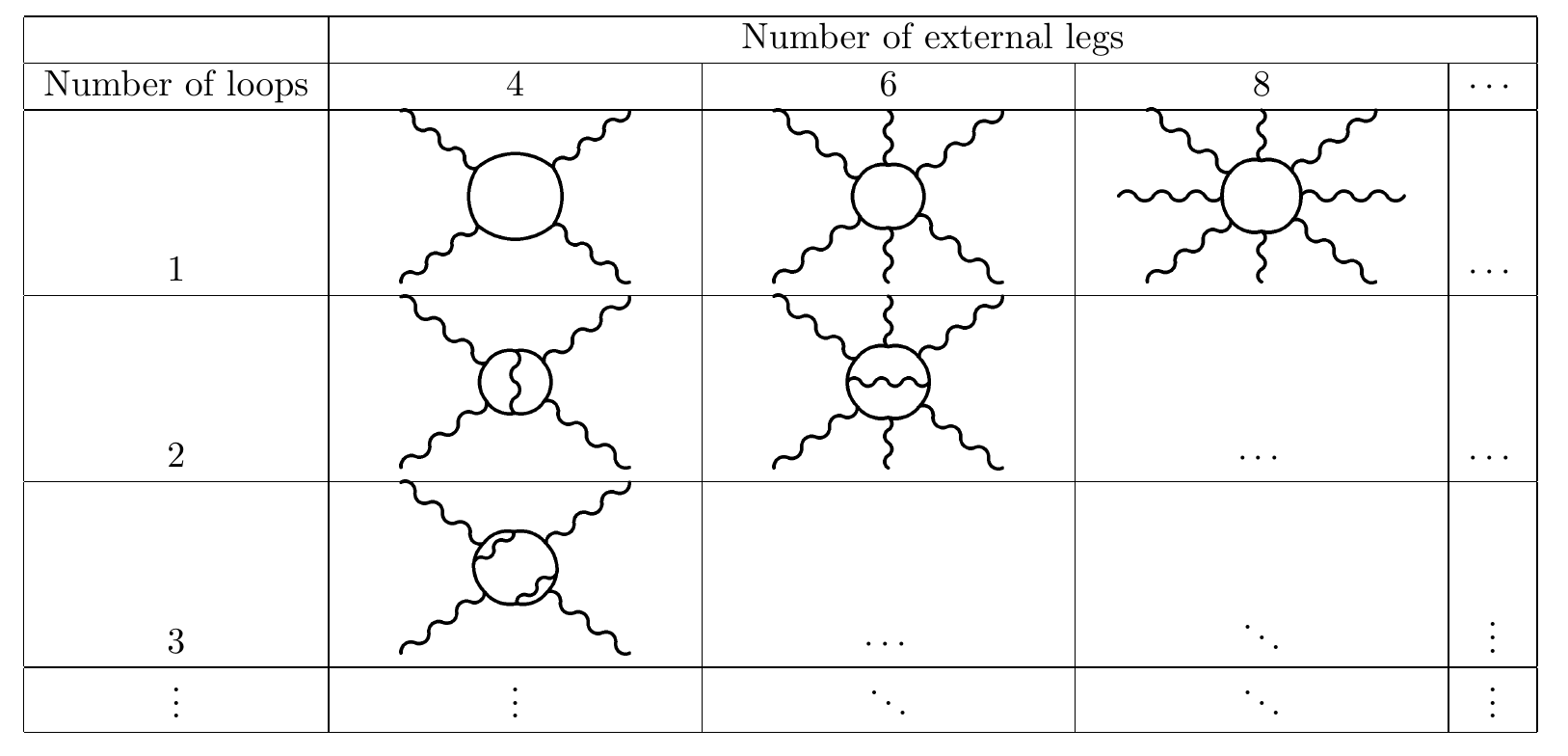}
}
\caption{Feynman diagrams contributing to the AAM formula.}
\label{AAMfeyn}
\end{figure}

Thus Affleck et al. arrive, with very little effort, at the same exponentiation for Scalar QED
that Lebedev and Ritus find in Spinor QED:

\bear
{\rm Im}{\cal L}^{(\rm all-loop)}_{\rm scal}(E)
 &=& \sum_{l=1}^{\infty}{\rm Im}{\cal L}^{(l)}_{\rm scal}(E)
 \nonumber\\
&{\stackrel{\beta\to 0}{\sim}}&
 -\frac{m^4\beta^2}{16\pi^3}
\,{\rm exp}\Bigl[ -{\pi\over\beta}+\alpha\pi \Bigr] \nonumber\\
&=&  {\rm Im}{\cal L}^{(1)}_{\rm scal}(E)\,\,{\rm e}^{\alpha\pi} \, .
\nonumber\\
\label{AAM}
\ear
Their argument assumes the field to be weak, but there is no restriction on the
strength of the coupling $\alpha$.  We note:

\begin{itemize}

\item
Formula (\ref{AAM}), if true, constitutes a rare case of an all-loop summation of an
infinite series of graphs of arbitrary loop order. Those graphs are shown in Fig. \ref{AAMfeyn}.

\item
According to \cite{afalma}, the contribution of all non-quenched diagrams gets suppressed in the weak-field limit. 

\item
Perhaps most surprisingly, the scalar mass appearing in (\ref{AAM}) is already {\it the physically renormalized one}, implying
that the worldline instanton approach automatically takes all mass renormalization counterdiagrams into account. This is remarkable
considering that the determination of the physical mass parameter for the EHL becomes a rather nontrivial issue already at two-loops
\cite{ritusspin,ditreubook,18}. 

\end{itemize}

\vspace{10pt}

Thus according to Affleck et al. a true all-loop summation has produced the factor ${\rm e}^{\alpha\pi}$, which is not only unreasonably simple,
but also perfectly analytical in the fine structure constant $\alpha$! 
According to what has been said above, this would seem to point towards extensive cancellations between Feynman diagrams.
However, neither \cite{afalma} nor \cite{lebrit} made this point; perhaps, because the Schwinger pair creation rate is a rather
peculiar quantity. Thus our next goal will be to transfer the exponential factor from the imaginary to the real part
of the EHL, by a Borel dispersion relation, and from there to the low-energy photon S-matrix through the same procedure as at one-loop:

\bear
{\rm Im}{\cal L}\quad \stackrel{\rm disp.\, rel.}{\longrightarrow}\quad  {\rm Re}{\cal L} 
\quad \longrightarrow \quad \Gamma[k_1,\varepsilon_1;\ldots;k_N,\varepsilon_N] \, .
\label{transferexpfact}
\ear

First we need to see whether our one-loop Borel dispersion relation can be extended to the multiloop level. 
For this it will be useful to consider the simplest possible nonzero constant field background, which is the self-dual one.

\section{The self-dual case}
\label{self-dual}
\renewcommand{\theequation}{5.\arabic{equation}}
\setcounter{equation}{0}

As we mentioned above, the two-loop correction to the EHL for a purely electric or purely magnetic field are known only in
terms of intractable integrals, and only the first few coefficients have been calculated so far.
However, this case is not the simplest one that one can consider; mathematically much better behaved is the one of a 
(Euclidean) constant  self-dual field, defined by
$F_{\mu\nu} = \half \varepsilon_{\mu\nu\alpha\beta}F^{\alpha\beta}$. The field strength tensor can be written as

\bear
F=\biggl( \begin{smallmatrix} 
0&f&0&0\\
-f&0&0&0\\
0&0&0&f\\
0&0&-f&0\\
\end{smallmatrix} \biggr)
\, .
\label{defFSD}
\ear
At the one-loop level, the self-dual (``SD'') EHLs for Scalar and Spinor QED are special cases of (\ref{ehspin}), (\ref{ehscal}). 
Anticipating the result of the two-loop calculation below, it will be useful here to eliminate the proper-time integral, and 
perform a change of variables from $f$ to  $\kappa\equiv \frac{m^2}{2ef}$. This leads to \cite{51}

\bear
{\cal L}_{\rm scal}^{(1)(SD)}(\kappa) 
&=& {m^4\over (4\pi)^2}\frac{1}{\kappa^2}\left[-{1\over
12}\ln(\kappa) +\zeta'(-1)+\Xi(\kappa)\right]
\label{L1SD}
\ear 
where the function $\Xi(x)$ is defined as follows:

\bear
\Xi(x)\equiv \int_0^x dy\,\ln\Gamma(y)
-x\ln\Gamma(x)+{x^2\over 2}\ln(x)-{x^2\over 4}-{x\over 2}
\, .
\label{defXi}
\ear
The spinor EHL in this SD case after renormalization differs from the scalar one only by a trivial global factor of $-2$
(the reason for the independence of spin is that the Dirac equation in such a background possesses a hidden supersymmetry \cite{dadvec}). 

Remarkably, for the SD case it is possible to do all integrals in closed form not only at one-loop,
but even at two-loop, in both Scalar and Spinor QED.
The results can be written compactly in terms of the digamma function $\psi(x)\equiv \Gamma'(x)/\Gamma(x)$: 

\bear
{\cal L}^{(2)}(f)
&=&
-2\alpha \,{m^4\over (4\pi)^3}\frac{1}{\kappa^2}\left[
3\xi^2 (\kappa)
-\xi'(\kappa)\right] \, ,
\nonumber\\
{\cal L}_{\rm scal}^{(2)}(f)
&=&
\alpha \,{m^4\over (4\pi)^3}\frac{1}{\kappa^2}\left[
{3\over 2}\xi^2 (\kappa)
-\xi'(\kappa)\right] \, .
\nonumber\\
\label{L2SD}
\ear
Here $\kappa\equiv \frac{m^2}{2ef}$ and 

\bear
\xi(x)\equiv -x\Bigl(\psi(x)-\ln(x)+{1\over 2x}\Bigr)
\label{defxi}
\ear
(note that $\xi(x) = \Xi'(x)$).
Using the well-known expansion of the digamma function at $x=\infty$ in terms of
the Bernoulli numbers, 

\begin{eqnarray}
\psi(x)&\sim& \ln x-\frac{1}{2x}-\sum_{k=1}^\infty \frac{B_{2k}}{2k
\,x^{2k}} 
\label{xiexpinfty}
\end{eqnarray}
one finds the following closed-form expressions for the one- and two-loop weak-field expansion coefficients $c_n^{(1,2)(SD)}$
(we write them down for the spinor case):

\bear
c^{(1)(SD)}_n&=& 
- \frac{{B}_{2n}}{2n(2n-2)} \, ,
\nonumber\\
c^{(2)(SD)}_n &=&
{1\over (2\pi)^2}\biggl\lbrace
\frac{2n-3}{2n-2}\,{B}_{2n-2}
+3\sum_{k=1}^{n-1}
{B_{2k}\over 2k}
{B_{2n-2k}\over (2n-2k)}
\biggr\rbrace
\, .
\nonumber\\
\label{SDcoeff}
\ear
Further, in this self-dual case there is also an analogue of the distinction between
a purely magnetic and a purely electric field. For $f$ real the SD EHL
turns out to have a weak-field expansion with alternating coefficients, so that it is
Borel summable, and there is no imaginary part. Thus we call this case ``magnetic-like''. 
Taking $f$ imaginary removes the alternating sign and creates a pole in the Borel integral,
which implies an imaginary part for the EHL. Thus we call this case ``electric-like''. 
This imaginary part of the self-dual EHL with 
complex $f$ is obtained from (\ref{L2SD}) simply by using the analytic continuation of the
digamma function, and thus also known in closed form. 

Studying the self-dual case turned out to be useful in three ways: 

\begin{itemize}

\item
For this case we could verify that the Borel dispersion relation (\ref{boreldisp}) can be used
to construct the imaginary part of the EHL from the weak-field expansion even at the two-loop
level. That is, the asymptotic three-parameter matching (\ref{fit}) works, and (more nontrivially)
the Borel summation procedure does not miss any non-perturbative terms (see \cite{dasdun} for a
case where such a thing actually occurred, even at one-loop).    

\item
The AAM exponentiation formula (\ref{AAM}) can be generalized to the SD case unchanged, by a simple
modification of the worldline instanton to a double circle,

\bear
x_{\rm instanton}(\tau) = {m\over \sqrt{2} eE}\bigr({\rm cos}(2\pi \tau/T),{\rm sin}(2\pi \tau/T),{\rm cos}(2\pi \tau/T),{\rm sin}(2\pi \tau/T)\bigl) 
\, .
\nonumber\\
\label{instantonSD}
\ear
And the initial step of the exponentiation, eq. (\ref{Im1plus2}),  is easy to verify 
explicitly from (\ref{L1SD}), (\ref{L2SD}). 
This holds independently of spin.
 
\item
The effective action for a self-dual field is unphysical, since such a field cannot be real
in Minkowski space. Nevertheless, it still carries information on the physical photon amplitudes;
the self-duality condition corresponds precisely to a projection on the 
`all $+$'  (or `all $-$') photon amplitudes \cite{dufish}. Thanks to the closed-form expressions (\ref{L2SD}),
even at the two-loop level we are still able to write down a closed-form all-$N$ expression for this
particular polarization choice:

\end{itemize}

\bear
\Gamma^{(1)}
[k_1,\varepsilon_1^+;\ldots ;k_N,\varepsilon_N^+]
&=& 
-2\frac{(2e)^{N}}{(4\pi)^2m^{2N-4}}\,c^{(1)(SD)}_{\scriptstyle\frac{N}{2}}
\chi_N \, ,\nonumber\\
\Gamma^{(2)}
[k_1,\varepsilon_1^+;\ldots ;k_N,\varepsilon_N^+]
&=&
-2\alpha\pi\frac{(2e)^{N}}{(4\pi)^2m^{2N-4}}\,c^{(2)(SD)}
_{\scriptstyle{\frac{N}{2}}}
\chi_N 
\, .
\nonumber\\
\label{GammaMHV}
\ear
As was mentioned above, here all the dependence on momenta and polarizations is carried by a unique
(independent of loop order) invariant $\chi_N$. Using spinor helicity techniques, this invariant can be constructed explicitly for all $N$ \cite{56}.

\section{Synthesis: a conjecture for the photon S matrix}
\label{conjecture}
\renewcommand{\theequation}{6.\arabic{equation}}
\setcounter{equation}{0}

We are now ready to state a conjecture for 
the $N$ photon amplitudes at arbitrary loop level $l$ \cite{60}.
Consider the $l$ - loop correction to the purely electric EHL,
and define its weak-field expansion coefficients by

\bear
{\cal L}^{(l)}(E) = \sum_{n=2}^{\infty} c^{(l)}(n) \Bigl(\frac{eE}{m^2}\Bigr)^{2n}
\label{defcn}
\ear
(note the change of convention with respect to (\ref{EHLmag})).
Assuming that the AAM formula (\ref{AAM}) holds, and that the Borel dispersion relation (\ref{boreldisp})
works at each loop order, we can conclude that the leading asymptotic factorial growth rate must be the same at each loop 
order, namely $\sim \Gamma(2n-2)$:

\bear
c^{(l)}(n)\,\, {\stackrel{n\to \infty}{\sim}} \,\,c^{(l)}_{\infty}\, \pi^{-2n}\Gamma(2n - 2) \, .
\label{cnleading}
\ear
And here $c^{(l)}_{\infty}$ relates to the leading Schwinger exponential at $l$ loops,

\bear
{\rm Im}{\cal L}^{(l)}(E) {\stackrel{\beta\to 0}{\sim}} c^{(l)}_{\infty}\,\e^{-\frac{\pi m^2}{eE}} \, .
\label{displ}
\ear
At two loops, the numerical calculations of \cite{37} confirm this, but  only if physical mass renormalization is used!
For generic mass renormalization one finds instead a leading factorial behavior of $\Gamma(2n)$, and it is
only through a cancellation of this leading order term between the unrenormalized EHL and its mass renormalization counter term
that this leading factorial behavior gets reduced to the same $\Gamma(2n-2)$ behavior as at one loop. At the $l$ - loop level, it is still
not difficult to establish the leading factorial growth of the weak field expansion coefficients before renormalization, which is  

 \bear
 c^{(l)}(n)\,\, {\stackrel{n\to \infty}{\sim}} \,\,\Gamma(2n+2l-4) \, .
 \label{leadingunren}
 \ear
Thus at higher loop orders the correctness of the AAM conjecture requires increasingly 
extensive cancellations in the mass renormalization process to cut the
leading factorial growth all the way down to $\Gamma(2n - 2)$. 

Now let us consider the ratio of the $l$ - loop to the one-loop coefficients.
Combining (\ref{cnleading}), (\ref{displ}) with the AAM conjecture (\ref{AAM}), we find
at any fixed loop order 

\bear
{{\rm lim}_{n\to\infty}} {c^{(l)}(n)\over c^{(1)}(n)} \, = \,
 {c^{(l)}_{\infty}\over c^{(1)}_{\infty}}
 \quad \stackrel{AAM}{ =} \quad
{1\over (l-1)!}(\alpha\pi)^{l-1} \, .
\nonumber\\
\ear
At this stage, let us switch to the self-dual case. This is not essential for our argumentation,
but we prefer it for two reasons: First, more is known explicitly about the SD EHL; second, as mentioned above
the SD EHL directly translates into one particular helicity component
of the $N$ - photon amplitude, the one with all helicities equal (``all +'' or ``all  -''). 

Using our above rule for the conversion of the self-dual weak-field expansion coefficients
into the ``all +'' photon amplitudes, which is independent of the loop order, 
we get the following statement for the ``all +'' amplitudes in the limit of large photon number $N=2n$,

\bear
{\rm lim}_{N\to\infty}{\Gamma^{(l)}
[k_1,\varepsilon_1^+;\ldots ;k_N,\varepsilon_N^+]
\over
\Gamma^{(1)}
[k_1,\varepsilon_1^+;\ldots ;k_N,\varepsilon_N^+]
} = 
{{\rm lim}_{N\to\infty}} {\chi_N c^{(l)}(N/2)\over\chi_N c^{(1)}(N/2)} 
=
{1\over (l-1)!}(\alpha\pi)^{l-1} \, .
\nonumber\\
\label{limratio}
\ear
Summing this relation over $l$ we get

\begin{figure}[htbp]
\begin{center}
\includegraphics{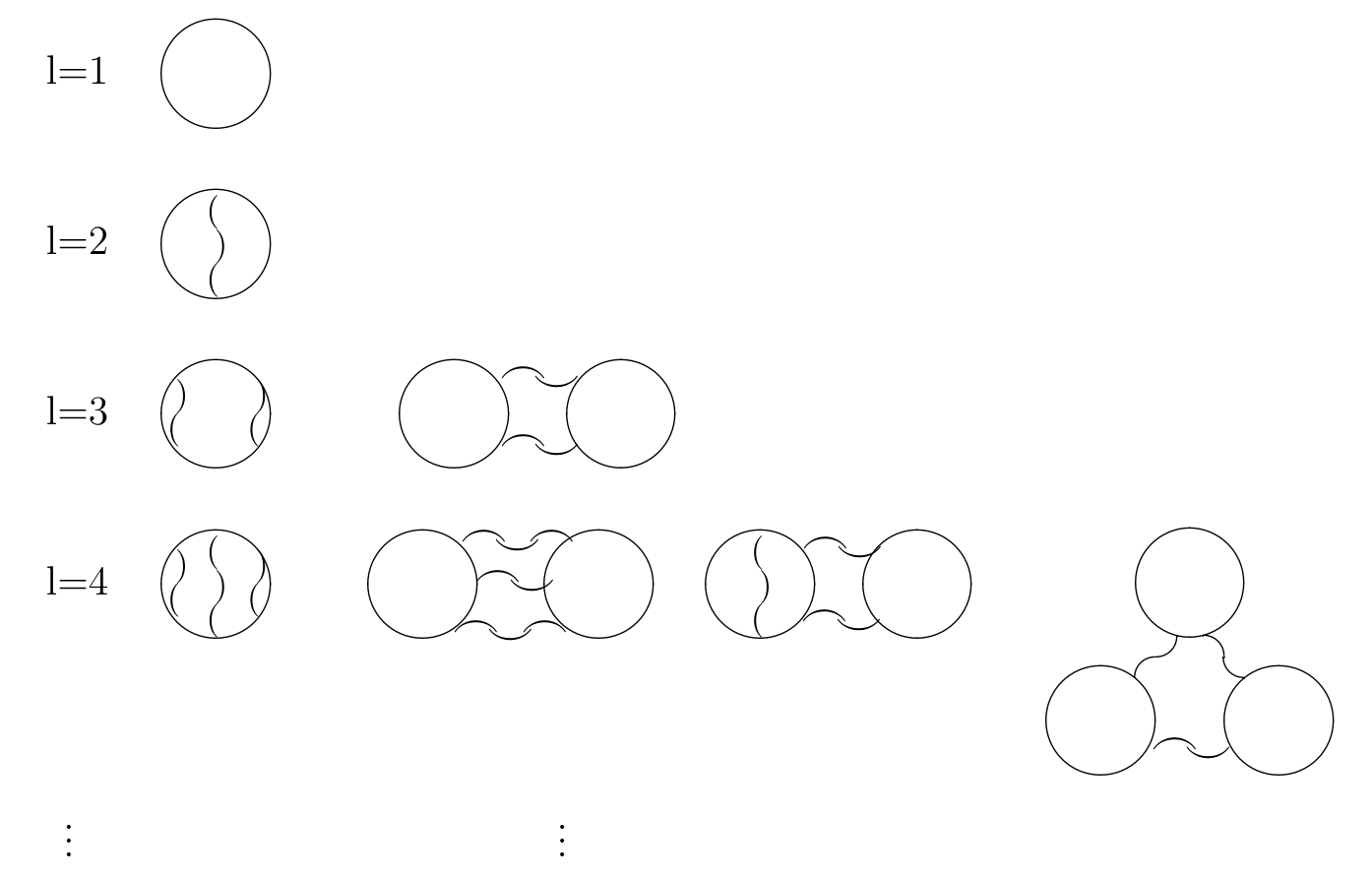}
\caption{Diagrams contributing to the EHL up to four loops (external legs not shown).}
\label{fig-colima}
\end{center}
\end{figure}

\bear
{\rm lim}_{N\to\infty}{\Gamma^{({\rm total})}
[k_1,\varepsilon_1^+;\ldots ;k_N,\varepsilon_N^+]
\over 
\Gamma^{(1)}
[k_1,\varepsilon_1^+;\ldots ;k_N,\varepsilon_N^+]}
=
\e^{\alpha\pi}
\, .
\label{conj}
\ear
Assuming sufficient uniformity in $l$ 
of the convergence for $N\to\infty$, one could now conclude that the amplitude must be analytic in $\alpha$ for some
sufficiently large $N$. But analyticity of the complete amplitude can certainly be safely excluded by renormalons and 
other arguments. Therefore uniformity must fail, and it is easy to see how this comes about 
diagramatically. In Fig. \ref{fig-colima} we show the diagrams contributing to the EHL up to four loops, not showing the external legs.

In the worldline-instanton based derivation of the AAM conjecture (\ref{AAM}) , only quenched diagrams contribute to the weak-field
limit of the imaginary part of the electric EHL, thus this must also be true for the leading asymptotic terms in the large $N$ expansion
of the weak-field expansion coefficients; non-quenched diagrams must get suppressed for $N\to\infty$. However, the
number of such diagrams is strongly growing with the loop order, so that  the process of the suppression of the non-quenched contributions
by increasing $N$ should slow down with increasing $l$. This provides a good reason for uniformity to fail for the whole amplitude, 
but there is no obvious reason to expect such a non-uniformity if one stays inside the class of quenched diagrams from the beginning. 
This led G.V. Dunne and one of the authors in 2004 \cite{60} to conjecture that {\it perturbation theory converges for the QED photon amplitudes
in the one electron-loop approximation}.
If true, this would imply enormous cancellations between
Feynman diagrams, presumably due to gauge invariance.

Only afterwards we learned that, as mentioned in the beginning, Cvitanovic \cite{cvitanovic77} had conjectured the analogous statement for the electron $g-2$ factor.

\section{Three predictions for the three-loop EHL}
\label{predictions}
\renewcommand{\theequation}{7.\arabic{equation}}
\setcounter{equation}{0}

To either disprove or further corroborate this conjecture, a calculation of the EHL at the three-loop level is called for.
We would like to see the following three things happen:

\benn

\item
We should see the next term of the exponentiation:

\bear
 \lim_{n\to\infty} \frac{c^{(3)}(n)}{c^{(1)}(n)} = \half (\alpha\pi)^2 \, .
 \label{alpha2}
 \ear

\item
At three loops there is already a non-quenched contribution, and it should 
be suppressed in the large $N$ limit.

\item

The convergence of 
${\frac{c^{(3)}(n)}{c^{(1)}(n)}}$
should not be slower than the one of
${\frac{c^{(2)}(n)}{c^{(1)}(n)}}$ 
when only quenched diagrams are taken.

\enn

\noindent
However, a calculation of the three-loop EHL in $D=4$ seems presently technically out of reach.

\section{ QED in 1+1 dimensions}
\label{1+1}
\renewcommand{\theequation}{8.\arabic{equation}}
\setcounter{equation}{0}

The proper-time representation of the one-loop EHL is essentially independent of dimension.
In 2006 M. Krasnansky \cite{krasnansky} studied the Scalar EHL in various dimensions also at two loops, and found, in particular,
the rather surprising fact that the Scalar EHL in 1 + 1 dimensions even at two-loop has a structure almost identical to the one of the self-dual Scalar EHL in
3+1 dimensions. Let us contrast the two cases: above we wrote down the self-dual field strength tensor for $D=4$,

\bear
F=\biggl( \begin{smallmatrix} 
0&f&0&0\\
-f&0&0&0\\
0&0&0&f\\
0&0&-f&0\\
\end{smallmatrix} \biggr) \, .
\nonumber
\ear
We also gave the Scalar EHL for this background,

\bear
{\cal L}_{\rm scal}^{(2)(4D)}(\kappa)
&=&
\alpha \,{m^4\over (4\pi)^3}\frac{1}{\kappa^2}\left[
{3\over 2}\xi^2 
-\xi'\right]\, ,\nonumber\\
\xi(\kappa) &=& -\kappa\Bigl(\psi(\kappa)-\ln(\kappa)+\frac{1}{2\kappa}\Bigr) \, .
\nonumber
\ear
In 2D the field strength tensor is
$F=\bigl( \begin{smallmatrix}   0 & f\\-f & 0 \end{smallmatrix} \bigr)$, and the
two-loop Scalar QED EHL comes out as \cite{krasnansky}

\bear
{\cal L}_{\rm scal}^{(2)(2D)}(\kappa)
&=&
-\frac{e^2}{32\pi^2}\left[
\xi^2_{2D} 
-4\kappa \xi_{2D}'\right] ,
\nonumber\\
\xi_{2D} &=& -\Bigl(\psi(\kappa+\half)-\ln (\kappa)\Bigr) \, .
\nonumber\\
\label{L2scal}
\ear
Since higher-loop calculations are somewhat easier in two dimensions,
this suggests to use the 2D case as a toy model for studying the AAM
conjecture. An effort along these lines was started in \cite{81}, however switching from
Scalar to Spinor QED.
Here we derived an analogue of the AAM conjecture in 2D, also using the
worldline instanton approach, and established the correspondences between
the 4D and 2D cases shown in Table 1:

\bear
4D\,\, QED \qquad &\leftrightarrow & \qquad 2D \,\, QED  \nonumber\\
 \alpha = \frac{e^2}{4\pi}
&\leftrightarrow &\qquad 
\tilde\alpha = \frac{2e^2}{\pi m^2} 
\nonumber\\
{\rm Im}\Gamma^{D=4}
\sim \,{\rm e}^{-\frac{m^2\pi}{eE} +\alpha\pi }
\qquad
& \leftrightarrow &
\qquad
{\rm Im}\Gamma^{D=2}
\sim
\e^{-\frac{m^2\pi}{eE} + \tilde\alpha \pi^2  \kappa^2}
\nonumber\\
{{\rm lim}_{n\to\infty}} {c^{(l)}_{4D}(n)\over c_{4D}^{(1)}(n)} 
= {(\alpha\pi)^{l-1}\over (l-1)!}
\qquad\qquad
& \leftrightarrow &
\qquad
{{\rm lim}_{n\to\infty}} {c^{(l)}_{2D}(n)\over c^{(1)}_{2D}(n+l-1)} 
= {(\tilde\alpha\pi^2)^{l-1}\over (l-1)!} \nonumber\\
{\rm Mass\, \,renormalization\,\, essential} \quad & \leftrightarrow &\quad
{\rm Mass\,\, renormalization\,\, irrelevant}
\nonumber
\label{dictionary}
\ear

\centerline{Table 1: Correspondences between the 4D and 2D cases.} 

\bigskip

There are two essential differences. First, in 2D the fine structure constant $\tilde\alpha$ is
not dimensionless. Thus the exponent of the AAM formula (rhs of third line) here involves also a factor of
$\kappa^2$, which in the formula for the asymptotic behavior of the weak-field expansion coefficients
(rhs of fourth line) leads to a shift in the argument between the $l$ - loop and the one-loop coefficients.
Thus in 2D the leading asymptotic growth of the coefficients increases with increasing loop order,
as it does in the 4D case {\it before} mass renormalization, and correspondingly it can be shown that the
contributions to the EHL from mass renormalization are asymptotically subleading, and thus irrelevant
for our purposes (although the fermion propagator in 2D does not have UV divergences, mass renormalization 
is still a quite nontrivial issue, see \cite{94}  and refs. therein). Presumably this relates to the fact that QED in 2D
is confining. 

In any case, all three of our three-loop predictions above have an analogue in the 2D case.
In \cite{81} we also obtained the following formulas for the one- and two-loop EHLs in 2D Spinor QED:

\bear
{\cal L}^{(1)}(\kappa ) &=& -{m^2\over 4\pi} {1\over\kappa}
\Bigl[{\rm ln}\Gamma(\kappa) - \kappa(\ln \kappa -1) +
\half \ln \bigl({\kappa\over 2\pi}\bigr)\Bigr] \, ,
\nonumber\\
{\cal L}^{(2)}(f) &=& {m^2\over 4\pi}\frac{\tilde\alpha}{4}
\Bigl[ \tilde\psi(\kappa) + \kappa \tilde\psi'(\kappa)
+\ln(\lambda_0 m^2) + \gamma + 2 \Bigr] \, .
\nonumber\\
\label{EHL2D12loop}
\ear 
where we have now abbreviated

\bear
\tilde\psi (x) &\equiv& \psi(x) - \ln x + {1\over 2x} \, .
\nonumber
\ear
Comparing with the Scalar QED result (\ref{L2scal}), we see that 
the spinor QED one is significantly simpler, as it involves the digamma
function only linearly. This is another surprise, since in 4D the Scalar and Spinor
EHLs do not show structural differences.

Remarkably, the two-loop EHL can (up to an irrelevant constant) even be written in terms of derivatives 
of the one-loop EHL:

\bear
{\cal L}^{(2)}(f) &=& - \frac{\tilde\alpha}{4}\Bigl(m^2\partder{}{m^2}\Bigr)^2 {\cal L}^{(1)}(f) \, .
\ear
From (\ref{EHL2D12loop}) we find for the one- and two-loop weak-field expansion coefficients

\bear
c_{2D}^{(1)}(n) &=& (-1)^{n+1} \frac{B_{2n}}{4n(2n-1)} \, , \nonumber\\
c_{2D}^{(2)}(n) &=& (-1)^{n+1} \frac{\tilde\alpha}{8}\frac{2n-1}{2n}B_{2n} \, .\nonumber
\ear
From this we can,
using properties of the Bernoulli numbers, easily show that

\bear
\lim_{n\to\infty}  {c_{2D}^{(2)}(n)\over c_{2D}^{(1)}(n+1)} 
&=&
\tilde\alpha \pi^2 \, .
\nonumber
\ear
This verifies the 2D AAM-like formula of Table 1 at the linearized level.

\begin{figure}[h]
{\centering
\includegraphics{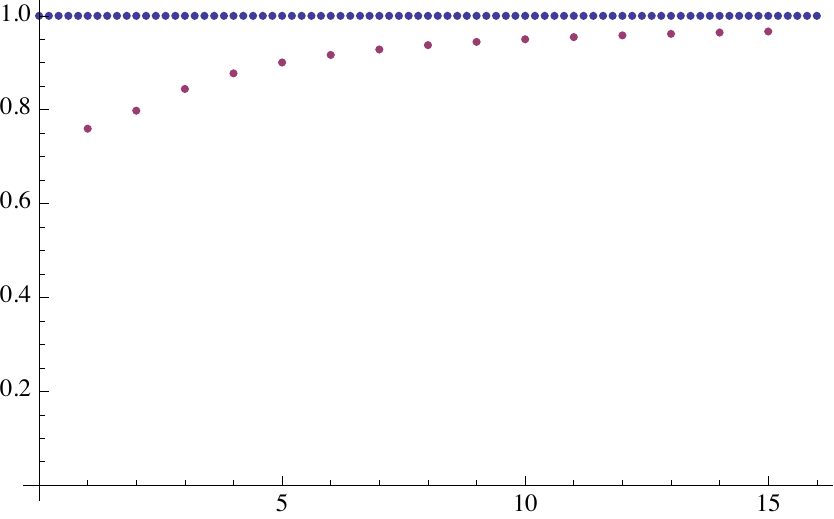}
}
\caption{Convergence of ${c_{2D}^{(2)}(n)\over c_{2D}^{(1)}(n+1)}$ to the AAM prediction (normalized such that the limit is unity) .}
\label{fig-ratio2c1}
\end{figure}

In Fig. \ref{fig-ratio2c1} we show the convergence to the asymptotic limit, which is rather rapid.

Even in the 2D case, the calculation of the three-loop EHL turned out be a formidable task, and it is only very recently that we were able
to obtain it in a form suitable for computing a sufficient number of the weak-field expansion coefficients \cite{85,inprep}.

\begin{figure}[h]
{\centering
\includegraphics{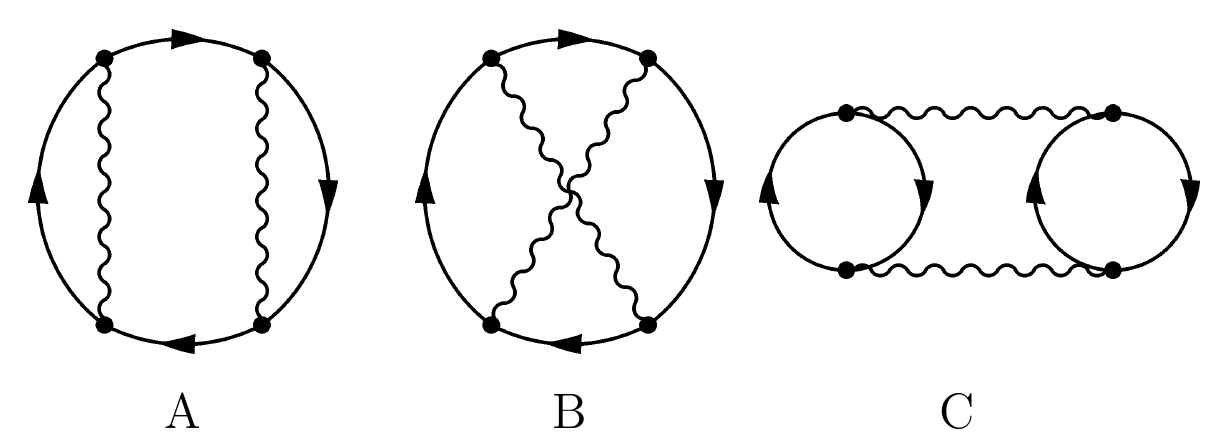}
}
\caption{Three-loop Feynman diagrams.}
\label{fig-ABC}
\end{figure}

At three loops, we have the three Feynman diagrams shown in Fig. (\ref{fig-ABC}).

Here the solid lines denote the electron propagator in the constant field. 
A and B are quenched, C is non-quenched.

\begin{figure}[h]
\centerline{\centering
\includegraphics[scale=.6]{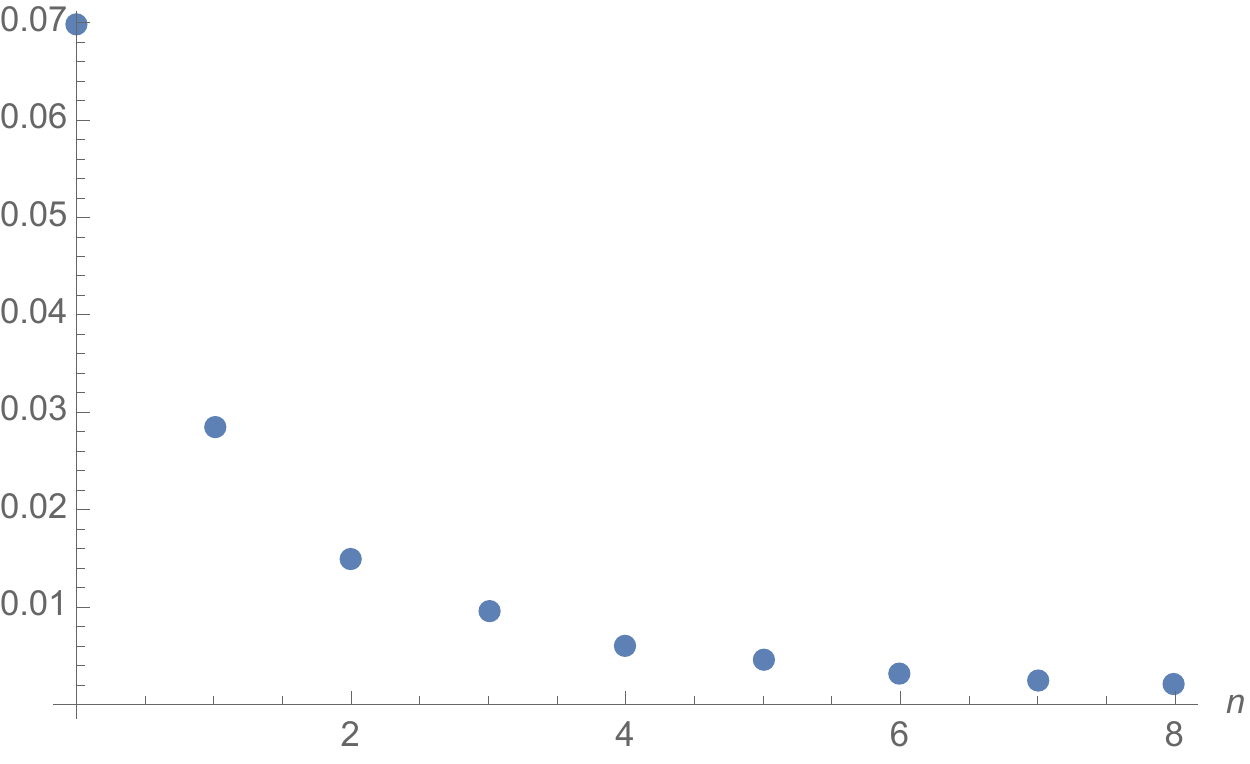}
}
\caption{The ratio ${c_{C}^{(3)}(n)\over c_{2D}^{(1)}(n+2)}/{(\tilde\alpha\pi^2)^2\over 2}$ for $n=0,\ldots,8$.}
\label{fig-diagCratio}
\end{figure}

The last one is by far the easiest one, and it is straightforward to obtain for it the following compact
integral representation:

\bear
\mathcal{L}^{3C}(f) &=& \frac{e^3}{16\pi^3 f}\int_{0}^{\infty} dz dz' d\hat{z} dz'' \frac{\sinh z \sinh z' \sinh \hat{z} \sinh z''}{[\sinh (z+z') \sinh (\hat{z}+z'')]^2}\nonumber  \\
& \times & \frac{\e^{-2\kappa (z+z'+\hat{z}+z'')}}{\sinh z \sinh z' \sinh (\hat{z}+z'') +\sinh \hat{z} \sinh z'' \sinh (z+z')} \, . \nonumber\\
\label{diagCrep}
\ear
This representation turned out to be quite adequate for a numerical 
calculation of the first 9 weak-field expansion coefficients $c_C^{(3)}(n)$ of diagram C.
In Fig. \ref{fig-diagCratio} we use these nine coefficients to show that
this unquenched contribution is indeed {asymptotically subleading}.

\noindent
This settles point 2. of our wish list above!

\begin{figure}[h]
\centerline{\centering
\includegraphics[scale=.6]{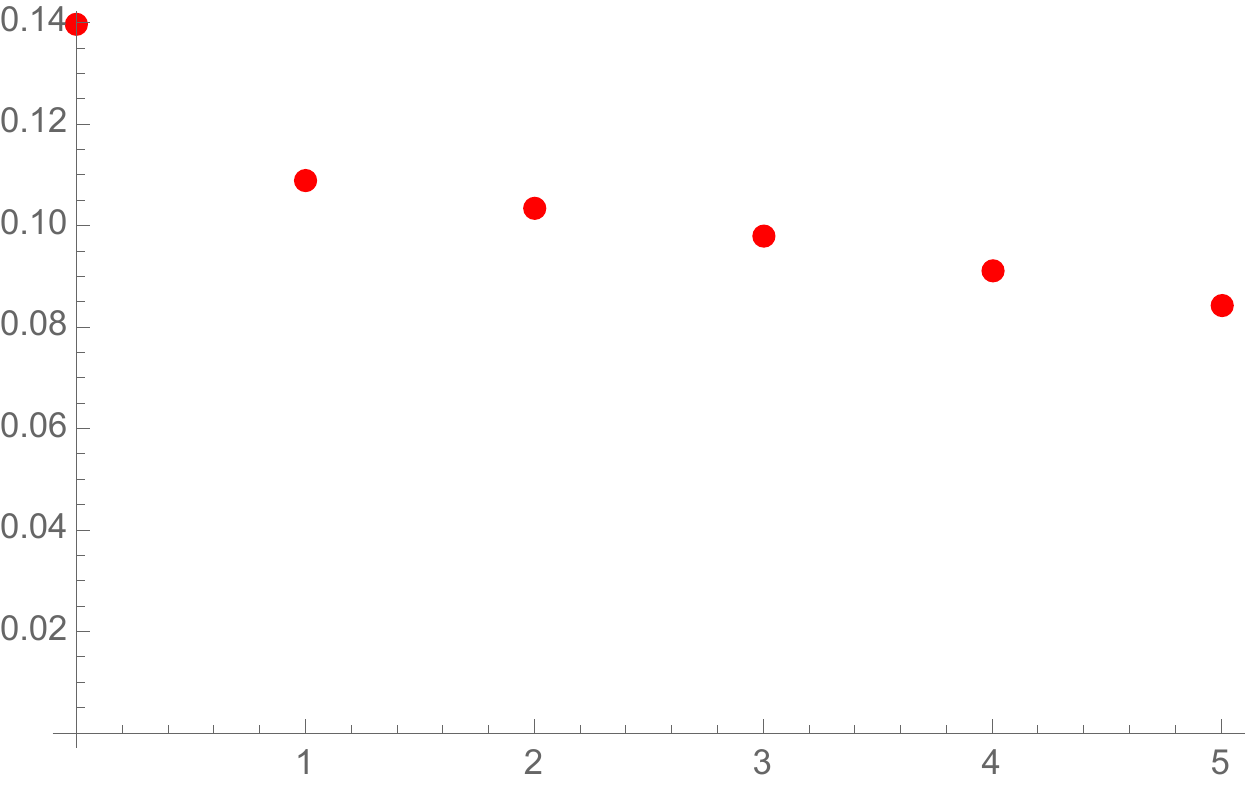}
}
\caption{The ratio ${c_{A+B}^{(3)}(n)\over c_{2D}^{(1)}(n+2)}/{(\tilde\alpha\pi^2)^2\over 2}$ for $n=0,\ldots,5$.}
\label{fig-ratioc3c1}
\end{figure}

Diagrams A + B are much more difficult, but the use of the ``traceless
gauge'' choice $\lambda = 2$ led to simplifications, and in
particular to manifest IR finiteness term-by-term. 
We managed to compute the first coefficient analytically,

\bear
c^3_{A+B}(0) =  \Bigl(-\frac{3}{2} + \frac{7}{4}\zeta (3) \Bigr) \frac{\tilde\alpha^2}{64}
\label{c3AB}
\ear
and five more coefficients numerically. 
Using these to plot the ratio 
${c_{A+B}^{(3)}(n)\over c_{2D}^{(1)}(n+2)}/{(\tilde\alpha\pi^2)^2\over 2}$
we get Fig. \ref{fig-ratioc3c1}.

Thus we are falling even below the asymptotic prediction!

\section{Conclusions and Outlook}
\label{concout}
\renewcommand{\theequation}{9.\arabic{equation}}
\setcounter{equation}{0}

Let us summarize:

\begin{itemize}

\item
We have presented first results of a calculation of the three-loop 2D EHL. This is the first calculation of
a three-loop effective Lagrangian in QED. 

\item
Although so far we have been able to compute only six coefficients of the weak-field expansion
(we should be able to obtain a few more) it seems already likely that the analogue of the AAM conjecture fails in 2D QED. 
This would throw also serious doubts on the validity of the 4D AAM conjecture.

\item
However, since the coefficient ratios fall below, rather than above, the asymptotic prediction,
{\it the riddle of the unreasonable smallness of loop corrections remains}. Presumably the worldline
instanton approach captures some valid information on large-scale cancellations between Feynman
diagrams, but needs refinement beyond two loops.

\item
We have also made an effort to make the point that
the QED photon amplitudes in the limit of low energy and large number of photons are very natural
objects for a study of the asymptotic properties of the QED perturbation series.

\item
It should also have become clear that physical mass renormalization is essential for asymptotic estimates in QED!
Unless mass renormalization is done physically, QED perturbation theory will break down already
at the two-loop level, because the two-loop contribution to any helicity component of the
$N$ - photon amplitude will, at least in the low energy limit, dominate over the one-loop one for sufficiently large $N$.
This implies, in particular, that approaches to the study of the high-order behavior of the QED perturbation series that are indifferent
to the issue of physical mass renormalization ought to be viewed with great caution.

\item
As an aside, it would be interesting to study also the QCD $N$ - gluon amplitudes for large $N$ from the point of view
of mass renormalization. 

\end{itemize}

\noindent
The authors declare that there is no conflict of interest regarding the publication of this paper.

%
%


\end{document}